\documentclass[sigconf]{acmart}

\AtBeginDocument{}

\setcopyright{none}
\renewcommand\footnotetextcopyrightpermission[1]{} 
\begin{document}

\title{ACM Multimedia Grand Challenge on Detecting Cheapfakes}

\author{Shivangi Aneja}\affiliation{\country{}\institution{Technical University of Munich}}
\author{Cise Midoglu}\affiliation{\country{}\institution{SimulaMet}}
\author{Duc-Tien Dang-Nguyen}\affiliation{\country{}\institution{University of Bergen}\institution{Kristiania University College}}
\author{Sohail Ahmed Khan}\affiliation{\country{}\institution{University of Bergen}}
\author{Michael Riegler}\affiliation{\country{}\institution{SimulaMet}}
\author{Pål Halvorsen}\affiliation{\country{}\institution{SimulaMet}}
\author{Chris Bregler}\affiliation{\country{}\institution{Google AI}}
\author{Balu Adsumilli}\affiliation{\country{}\institution{YouTube}}

\renewcommand{\shortauthors}{Aneja et al.}

\begin{abstract}
\textit{Cheapfake} is a recently coined term that encompasses non-AI (``cheap'') manipulations of multimedia content. Cheapfakes are known to be more prevalent than deepfakes. Cheapfake media can be created using editing software for image/video manipulations, or even without using any software, by simply altering the context of an image/video by sharing the media alongside misleading claims. This alteration of context is referred to as out-of-context (OOC) misuse of media. OOC media is much harder to detect than fake media, since the images and videos are not tampered. In this challenge, we focus on detecting OOC images, and more specifically the misuse of real photographs with conflicting image captions in news items. The aim of this challenge is to develop and benchmark models that can be used to detect whether given samples (news image and associated captions) are OOC, based on the recently compiled COSMOS dataset.
\end{abstract}


\begin{teaserfigure}
  \includegraphics[width=\textwidth]{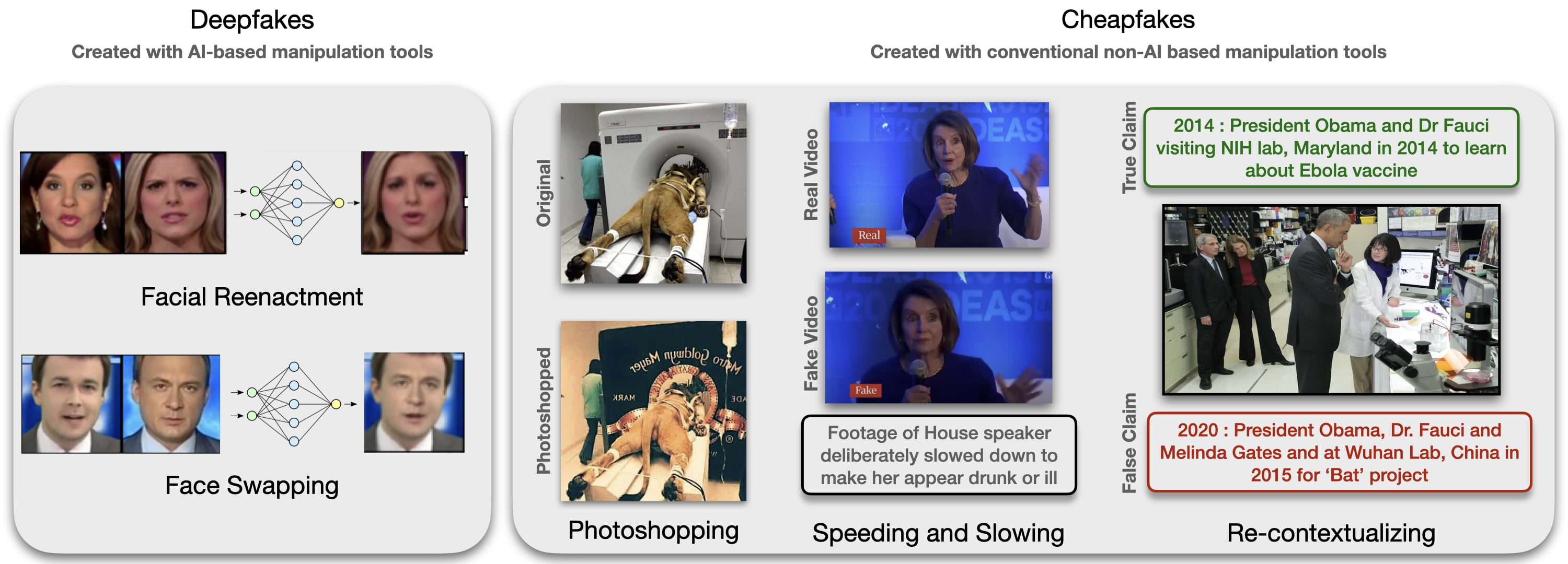}
  \caption{Deepfakes (left) are defined as falsified media created using sophisticated AI-based media manipulation tools and techniques. Cheapfakes (right) include falsified media created with/without contemporary non-AI based editing tools which are easily accessible. Photoshopping tools can be used to tamper with images. Videos can be sped up or slowed down to change the intent or misrepresent the person in the video. Re-contextualizing includes associating falsified or unrelated claims with a genuine image to misrepresent events or persons. This challenge is focused on detecting re-contextualized cheapfakes. Image sources:~\cite{ff_dataset,photoshopped_lion,pelosi_fake,obama_maryland,obama_wuhan}}
    \label{figure:teaser}
\end{teaserfigure}

\maketitle

\section{Introduction}\label{section:introduction}

The last decade has seen a surge in the use of social media platforms as a means of consuming news. In a recent study~\cite{forbes_report}, Forbes reports that social media giant Facebook leads this trend with $36\%$ of its customers using the platform for consuming news. Social media platforms come with a freedom for users to upload and share posts, which has led to the proliferation of fake media on these platforms.

\begin{figure*}
    \centering
    \includegraphics[width=0.7\linewidth]{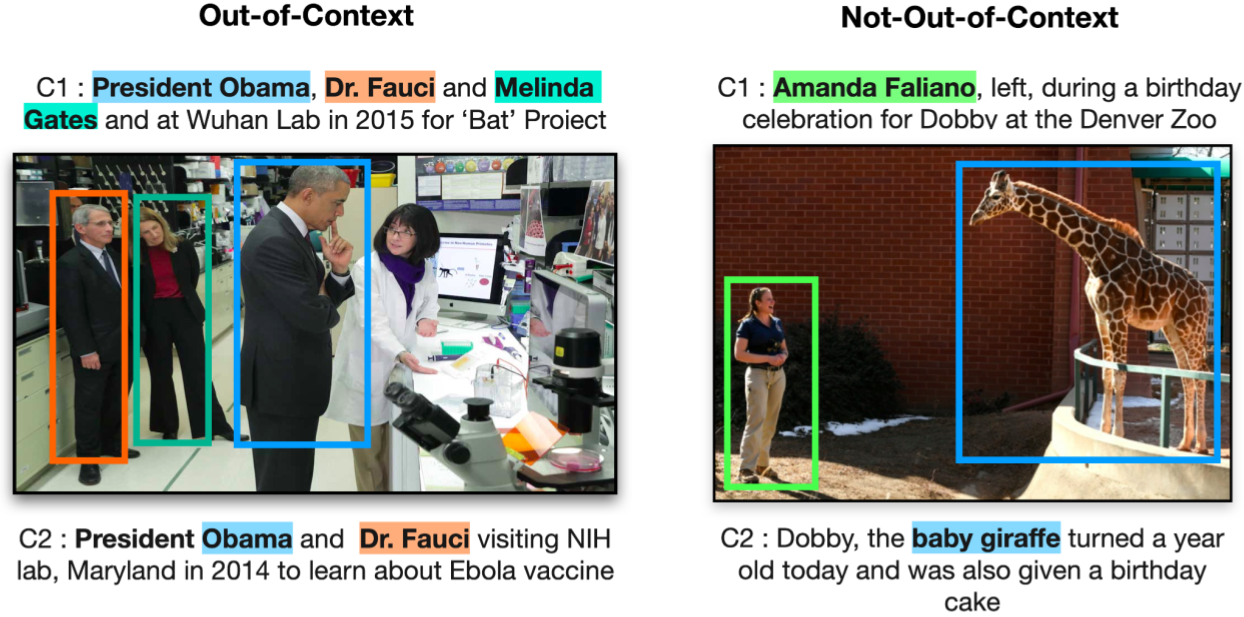}
    \caption{Each image in the dataset is accompanied by one or two captions that the image was circulated together with on the Internet. On the left, one of the two captions is misleading with an alteration of context, indicating out-of-context (OOC) misuse. On the right, none of the two captions are misleading, hence not-out-of-context (NOOC). Image source:~\cite{aneja2021cosmos}}
    \label{figure:task-figure}
\end{figure*}

Fake media (including audio, images, videos, and text) circulated on social media platforms can be broadly grouped into two major categories: \textit{deepfakes} and \textit{cheapfakes}, as shown in Figure~\ref{figure:teaser}. Deepfakes are falsified media, most commonly facial videos created using sophisticated AI-based media manipulation tools and techniques. Several deepfake detection methods~\cite{ff_dataset, Nguyen_2019, face_warping, deepfake_inconsistent_head_pose, cozzolino2018forensictransfer,mesonet, agarwal_protecting_2019, li2020face, verdoliva2020media, aneja2020generalized, cozzolino2020idreveal} are in place to monitor and regulate the spread of deepfake videos. \textit{Cheapfake} is a recently coined general term that encompasses non-AI (``cheap'') manipulations of multimedia content, created without using deep learning methods. Although a lot of attention has been paid to the creation, detection, and misuse of deepfakes in the last years, cheapfakes are actually known to be more prevalent than deepfakes~\cite{factsheet-covid19, mit_tech_report, reporterslab}. 

Cheapfakes are created with or without contemporary editing tools which are non-AI based and are easily accessible. Image manipulations, speeding/slowing of videos, and deliberate alteration of the context of the multimedia asset in, e.g., news captions, by sharing the media alongside misleading claims, are some of the methods that are currently in use (see Figure \ref{figure:teaser}). The latter is referred to as out-of-context (OOC) misuse of media. OOC media are much harder to detect than fake media, since the images and/or videos are not tampered. We refer readers to the report by Paris \textit{et al.}~\cite{paris2019} for an overview of different types of cheapfakes surfacing the Internet. 

Depending on the type of cheapfakes, different detection tools can be used. Methods to detect image manipulations such as photoshopping and image splicing have been investigated~\cite{Chen2017ImageSD, Cozzolino2015SplicebusterAN, huh2018fighting, wang2019detecting}. Re-contextualization or OOC misuse, which include associating falsified or unrelated claims with a genuine image in order to misrepresent events or persons is, however, relatively niche and unexplored. Very recently, Aneja et al.~\cite{aneja2021cosmos} introduced this task, provided a dataset of real-world news posts called COSMOS, and proposed a method for detecting cheapfakes, which was benchmarked using the COSMOS dataset. 

In this challenge, we focus on detecting OOC images, and more specifically the misuse of real photographs with conflicting image captions, in news items. The aim of this challenge is to develop and benchmark models that can be used to detect whether given samples (news image and associated captions) are OOC, based on a version of the COSMOS dataset.

\section{Challenge Tasks}

\subsection*{Task 1: Identification of Conflicting Image-Caption Triplets}

An image serves as evidence of the event described by a news caption. If two captions associated with an image are valid, then they should describe the same event. If they align with the same object(s) in the image, then they should be broadly conveying the same information. Based on these patterns, we define out-of-context (OOC) use of an image as presenting the image as an evidence of untrue and/or unrelated event(s). If the two captions refer to same object(s) in the image, but are semantically different, i.e., associate the same subject to different events, this indicates OOC use of the image. However, if the captions correspond to the same event, irrespective of the object(s) the captions describe, this is defined as not-out-of-context (NOOC) use of the image. See Figure~\ref{figure:task-figure} for more details. 

In this task, the participants are asked to come up with methods to detect conflicting image-caption triplets, which indicates miscontextualization. More specifically, given \texttt{<Image,Caption1,Caption2>} triplets as input, the proposed model should predict corresponding class labels (OOC or NOOC). The end goal for this task is not to identify which of the two captions is true/false, but rather to detect the existence of miscontextualization. This kind of a setup is considered particularly useful for assisting fact checkers, as highlighting conflicting image-caption triplets allows them to narrow down their search space.

\subsection*{Task 2: Fake Caption Detection}

A NOOC scenario from Task 1 makes no conclusions regarding the veracity of the statements. In a practical scenario, multiple captions might not be available for a given image. In such a scenario, the task boils down to figuring out whether a given caption linked to the image is genuine or not. We argue that this is a challenging task, even for human moderators, without prior knowledge about the image origin. Luo et al.~\cite{luo2021newsclippings} verified this claim with a study on human evaluators who were instructed not to use search engines, where the average human accuracy was around $65\%$. 

In this task, the participants are asked to come up with methods to determine whether a given \texttt{<Image,Caption>} pair is genuine (\textit{real}) or falsely generated (\textit{fake}). Since our dataset only contains real, non-photoshopped images, it is suitable for a practical use case and challenging at the same time.

\section{Dataset}

\begin{figure}[h!]
    \centering
    \includegraphics[width=\linewidth]{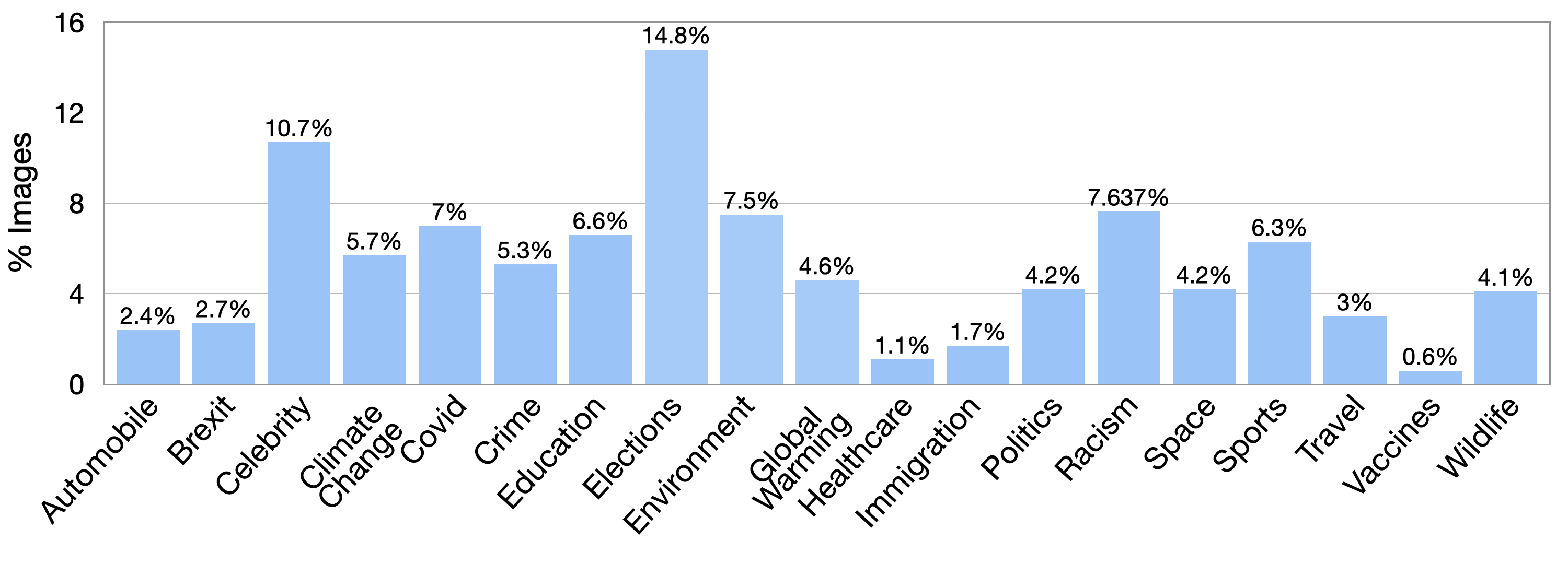}
    \caption{Distribution of the images in the COSMOS dataset per category.}
    \label{figure:dataset-distribution}
\end{figure}

Aneja et al.~\cite{aneja2021cosmos} have created COSMOS, a large-scale dataset of around $200K$ images which have been matched with $450K$ textual captions from different news websites, blogs, and social media posts. Figure~\ref{figure:dataset-distribution} presents the category distribution of the images in the dataset, which were collected from a wide variety of articles with a special focus on topics where misinformation spread is prominent. 

For this challenge, a part of the COSMOS dataset~\cite{aneja2021cosmos} is sampled and assigned as the public dataset. The public dataset, consisting of the training, validation and public test splits, is provided openly to participants for training and testing their algorithms. The remaining part of the COSMOS dataset is augmented with new samples and modified to create the hidden test splits, which are not made publicly available, and will be used by the challenge organizers to evaluate the submissions. Details of the challenge dataset are summarized in Table~\ref{table:dataset-statistics}.

\begin{table}[h!]
\small
    \centering
    \caption{Challenge dataset statistics.}
    \begin{tabular}{|c|c|r|r|p{1.5cm}|}
    \hline
    \textbf{Task} & \textbf{Split} & \textbf{\# Images} & \textbf{\# Captions} & \textbf{Context Annotation} \\
    \hline
    & Training & $161752$ & $360749$ & No \\
    \hline
    & Validation & $41006$ & $90036$ & No \\
    \hline
    Task 1 & Public Test & $1000$ & $2000$ & Yes\\
    \hline
    Task 2 & Public Test & $100$ & $100$ & Yes\\
    \hline
    \end{tabular}
    \label{table:dataset-statistics}
\end{table}

\subsection{Training and Validation Splits}

We provide common training and validation splits for both challenge tasks. The training and validation splits are provided as JavaScript Object Notation (JSON) formatted text files called \verb|train.json| and \verb|val.json|, where each data sample is stored as a dictionary~\footnote{https://github.com/shivangi-aneja/COSMOS}\label{footnote:cosmos}. The attributes in \verb|train.json| and \verb|val.json| are as follows.

\begin{itemize}
    \item \verb|img_local_path:| Source path for the image in the dataset directory.
    
    \item \verb|articles:| List of dictionaries containing metadata for every caption associated with the image.
    
    \item \verb|caption:| Original caption scraped from the news website.
    
    \item \verb|article_url:| Link to the website from where the image and caption were scraped.
    
    \item \verb|caption_modified:| Modified caption after applying Spacy NER\footnote{https://spacy.io/api/entityrecognizer}. Authors in~\cite{aneja2021cosmos} use this caption as an input to their model during experiments.
    
    \item \verb|entity_list:| List of mappings between the modified named entities in caption with the corresponding hypernyms.
    
    \item \verb|maskrcnn_bboxes:| List of detected bounding boxes corresponding to the image. (x1,y1) refers to the start vertex of the rectangle and (x2, y2) refers to end vertex of the rectangle. Note that for detecting bounding boxes, the authors in~\cite{aneja2021cosmos} use the Detectron2 pretrained model\footnote{https://github.com/facebookresearch/detectron2/blob/master/MODEL\_ZOO.md} available under\footnote{https://github.com/facebookresearch/detectron2/blob/master/configs/COCO-Keypoints/keypoint\_rcnn\_X\_101\_32x8d\_FPN\_3x.yaml}. They detect up to $10$ bounding boxes per image.

\end{itemize}

\subsection{Test Splits}

We provide separate test splits for evaluating the performance of the two tasks. 

\textbf{Task 1 test split:} The public test split is provided as a JSON formatted text file called \verb|test.json|, and has the structure described in\textsuperscript{\ref{footnote:cosmos}}. The hidden
test split is structurally identical to the public test split. The attributes in \verb|test.json| are as follows.

\begin{itemize}
    \item \verb|img_local_path:| Source path for the image in the dataset directory.
    
    \item \verb|caption1:| First caption associated with the image.
    
    \item \verb|caption1_modified:| Modified \verb|caption1| after applying Spacy NER.
    
    \item \verb|caption1_entities:| List of mappings between the modified named entities in \verb|caption1| with the corresponding hypernyms.
    
    \item \verb|caption2:| Second caption associated with the image.
    
    \item \verb|caption2_modified:| Modified \verb|caption2| after applying Spacy NER.
    
    \item \verb|caption2_entities:| List of mappings between the modified named entities in \verb|caption2| with the corresponding hypernyms.
    
    \item \verb|article_url:| Link to the website from where the image and caption were scraped.
    
    \item \verb|label:| Class label indicating whether the two captions are out-of-context with respect to the image, where $1$=out-of-context (OOC) and $0$=not-out-of-context (NOOC).
    
    \item \verb|maskrcnn_bboxes:| List of detected bounding boxes corresponding to the image. (x1,y1) refers to the start vertex of the rectangle and (x2, y2) refers to the end vertex of the rectangle.
\end{itemize}

\textbf{Task 2 test split:} The public test split is provided as a JSON formatted text file called \verb|task_2.json|. The hidden test split is structurally identical to the public test split. The attributes in \verb|task_2.json| are as follows.

\begin{itemize}
    \item \verb|img_local_path:| Source path for the image in the dataset directory.
    
    \item \verb|caption:| Text caption associated with the image.
    
    \item \verb|genuine:| Classification label, where 0=out-of-context (OOC) and 1=not-out-of-context (NOOC). 
\end{itemize}

\section{Evaluation Criteria}

There are two considerations in the fulfillment of the above described tasks. Participant models will be evaluated and ranked according to two aggregate scores representing these considerations, each composed of $5$ and $3$ metrics respectively. 

\textbf{Effectiveness:} The first goal is to achieve high detection performance. This speaks to \textit{effectiveness}, which will be evaluated based on accuracy, precision, recall, F1-score, and Matthews correlation coefficient (MCC). Participants are asked to calculate these $5$ metrics for their model and include the values in their submission.

\textbf{Efficiency:} In certain scenarios, having an idea about the potential misuse of images in real-time and with minimal resources can be more important than the detection performance itself. We take this aspect into consideration by introducing an additional goal: having low latency and low complexity. This speaks to \textit{efficiency}, which will be evaluated based on latency\footnote{Average, i.e., arithmetic mean of the runtime per sample, calculated over all samples in the public test split (ms).}, number of parameters\footnote{Number of trainable parameters in the model (million).}, and model size\footnote{Storage size for the model (MB).}. Participants are asked to calculate these $3$ metrics for their model and include the values in their submission.

\section{Administrative Details}

A challenge website\footnote{\url{https://detecting-cheapfakes.github.io/}} has been setup with a commitment to be maintained at least for the next $3$ years. Along with the official website, a Google Group\footnote{\url{https://groups.google.com/g/grandchallenge-cheapfakes}} and a GitHub repository\footnote{\url{https://github.com/detecting-cheapfakes/detecting-cheapfakes-code}} have been established to support prospective participants. Interested participants can find the previously asked questions and join interactive discussions on these platforms.

\section{Conclusion and Outlook}

The ACM Multimedia Grand Challenge on Detecting Cheapfakes addresses the relatively new but prominent problem of detecting cheapfake multimedia content in news. More specifically, it focuses on the OOC misuse of images in news items. As emphasized in Section~\ref{section:introduction}, this is a relatively novel area of research, in comparison to deepfakes, and is distinguishable from the wider field of fake news in the following way: the term ``fake news'' traditionally refers to the use of either fake multimedia content or false captions, whereas OOC misuse refers to the scenario where the multimedia content in the news item (images in this case) is decidedly not fake, and the captions may or may not be false. Rather, the misinformation results from the OOC combination of the two. 

With this challenge, we firstly aim to motivate researchers to develop different methods for addressing this particular problem, and to benchmark a number of proposed models for detecting the OOC misuse of real photographs in news items. Secondly, through the
dissection and analysis of the COSMOS dataset, we aim to encourage the in-depth understanding, and later generation of (further), supervised datasets for the above described challenges. Algorithmic benchmarking is an efficient approach to analyze the results from different detection methods, but on top of evaluating the models themselves, the comparison of different approaches can help identify the potentials as well as the shortcomings of existing open datasets. 

There is a growing interest from the scientific community towards addressing the problem of misinformation in general, and towards the detection of deepfakes in particular. We hope that this challenge can increase awareness regarding the prominence of
cheapfakes as well, and that in the future, the methods presented within this context can evolve into systems that support researchers, regulatory bodies, news consumers, and the general public in their search for a safe and truthful information ecosystem. 

The first edition of this challenge has been organized within the ACM Multimedia Systems (MMSys) Conference as the ``MMSys'21 Grand Challenge on Detecting Cheapfakes''~\cite{mmsys21-challenge-web,mmsys21-challenge-arxiv}. The challenge will be sustained over the next years, as we believe that it is specific enough to be relevant and timely, as well as generic enough to evolve over time according to different needs. For instance, with the rising importance of synthetic data for research and training purposes, a possible future task could be the \textit{generation} of fake captions.

\bibliographystyle{ACM-Reference-Format}
\bibliography{references}

\end{document}